\documentclass[letterpaper, a4paper]{amsart}
\usepackage{amsmath}
\usepackage{latexsym}
\usepackage[all]{xy}
\usepackage{color}

\include{amslatex}

\numberwithin{equation}{section}
\begin{document}
\begin{title}[Maximal acceleration in laser-plasma acceleration]
{Some consequences of theories with maximal acceleration in laser-plasma acceleration}
\end{title}
\maketitle
\begin{center}
\author{Ricardo Gallego Torrom\'e\footnote{Email: rigato39@gmail.com}}
\end{center}
\bigskip

\begin{center}
\address{Frankfurt Institute for Advanced Studies\\
Ruth-Moufang-Stra{\ss}e 1, 60438 Frankfurt, Germany}
\end{center}

\begin{abstract}
In this paper we consider classical electrodynamic theories with maximal acceleration and some of their phenomenological consequences for  laser-plasma acceleration. It is shown that in a recently proposed higher order jet theory of electrodynamics, the maximal effective acceleration reachable by a consistent bunch of point charged particles being accelerated by the wakefield is damped for bunches containing large number of charged particles. We argue that such prediction of the theory is falsifiable. In the case of Born-Infeld kinematics, laser-plasma acceleration phenomenology provides an upper bound for the Born-Infeld parameter $b$. Improvements in the beam qualities will imply stronger constraints on $b$.
\end{abstract}
\section{Introduction}
The possible existence of a maximal proper acceleration in Nature has been under investigation for a long time \cite{Caianiello,Brandt1983,Caldirola}. Recently, this conjecture has been boosted with renewed interest because its potential relation with string theory \cite{ParentaniPotting, BowickGiddins, FrolovSanchez 1991}, loop quantum gravity \cite{RovelliVidotto} and with modifications of classical electrodynamics \cite{Schuller,Ricardo 2017}. Indeed, the existence of a maximal proper acceleration implies the modifications of the theories of special and general relativity, it has relevant theoretical consequences for quantum field theory \cite{NesterenkoFeoliLambiaseScarpetta}, in modifications of the power radiation law in quantum electrodynamics \cite{Papini2015} and for the resolution of gravitational singularities \cite{CaianielloGasperiniScarpetta,RovelliVidotto}. However, although experimental and phenomenological proposals to measure the maximal acceleration have been discussed \cite{Friedman et al.,Potzel,Schuller 2}, the current accepted status is that only lower bounds for such hypothetical maximal proper acceleration have been achieved \cite{Ricardo Nicolini 2018}.

In this paper we investigate consequences arising in laser-plasma acceleration from theories of electrodynamics containing a maximal proper acceleration. This line of research is motivated by the following considerations. In situations where radiation reaction effects are of relevance, there must be a modification of the theory of special relativity. In particular, the so called {\it clock hypothesis} \cite{Einstein1922} should break down in such situations \cite{Mashhoon1990}. The clock hypothesis can be stated as {\it the equivalence of an accelerated coordinate system with a family of instantaneous inertial coordinate systems}. This equivalence in combination with other assumptions of the relativity theory imply that the class of geometric spacetime structures depend only on the position and the velocity tangent field of ideal test particles.  Usually, this is interpreted as the requirement that spacetime can be described conveniently by a Lorentzian structure.  If gravitational effects are disregarded, the spacetime structure is the Minkowski flat metric, but in the general case where gravity is taken into account, the metric structure is a Lorentzian structure subjected to dynamical field equations\footnote{Incidentally, the most general smooth structure compatible with all the assumptions of the theory of general relativity, including compatibility with the clock hypothesis and the weak equivalence principle, are on the class of {\it generalized Berwald spacetimes} \cite{Ricardo 2017b}. Lorentzian spacetimes, are a relevant class of Berwald spacetimes.}. However, it was pointed out by B. Mashhoon that the clock hypothesis is in general not applicable when the {\it intrinsic dimensions} of the observer are of the same size than the {\it scales of the interactions} \cite{Mashhoon1990}. This happens in classical electrodynamic dynamical systems when radiation reaction effects are of relevance. Conversely, the violation of the clock hypothesis leads to modifications of the classical spacetime Lorentzian structure.

On the other hand and despite that the problem of finding a consistent classical theory for radiation reaction of point charged particles has attracted much attention for a long time, the current classical standard theory contains pathological physical consequences \cite{Dirac}. In particular, the Lorentz-Dirac equation is plagued by run-away and pre-accelerated solutions. It is remarkable that despite Dirac's theory is based upon well established physical principles, it leads to a model of point charged particle which is physically unacceptable.

 Initially in the framework of classical field theory, a well known attempt to solve this situation is based upon the Landau-Lifshitz equation \cite{LandauLifshitz}. It was also argued that the Landau-Lifshitz equation is in fact a well-defined effective description of the dynamics of a point charged equation \cite{Spohn1, Spohn2}. However, it is un-clear whether the Landau-Lifshitz equation is free of all the problems that its parent equation (the Lorentz-Dirac equation) has, since the Landau-Lifshitz equation is obtained from the Lorentz-Dirac equation under some approximations.

In this context of lack of a satisfactory resolution for the classical motion of a point charged particle, the author has suggested a novel theory of classical electrodynamical systems, a proposal that was schematically discussed in \cite{Ricardo 2017} and more extensively developed in a series of notes in \cite{Ricardo2012}. The fundamental idea is to consider an extension of the notions of electromagnetic field and spacetime structure in such a way that both depend upon the details of how the classical fields are being probed by the interacting classical test particles. This goal is achieved by assuming that fields and the spacetime metric depend on higher order differential jets, that is, not only on the spacetime location of test particle, but also on its speed and acceleration \cite{Ricardo2012,Ricardo2015,Ricardo 2017}. Such considerations lead to a new dynamical model for the point charged particle, where the particle is described by a second order differential equation \cite{Ricardo 2017}. It can be argued that the new dynamical law contains a maximal acceleration and that the spacetime structure is described by a {\it metric of maximal acceleration}, a geometric structure that depends on the world line, the velocity tangent vector and the acceleration of the test particle being used to probe the spacetime structure or the electromagnetic field \cite{Ricardo2015}.

We will argue that the theory of higher order jet electrodynamics could become falsifiable in the context of laser-plasma wakefield acceleration, where the effective acceleration reached by the bunch of electrons is very large. We will show that such accelerations are currently close in magnitude to an estimate of the {\it effective maximal proper acceleration} reachable for the bunch of accelerated particles within the framework of higher order jet electrodynamics. We will describe this phenomenological consequence and compare it with the analogous effect in other alternative theories of classical electrodynamics  containing a maximal proper acceleration. In particular, we will also consider the case of Born-Infeld theory \cite{BornInfeld,Schuller}. An upper bound for the Born-Indeld parameter $b$ is obtained, of the same order of magnitude than the upper bound obtained from Thomas's precession \cite{Schuller 2}. Future improvements in laser-plasma acceleration technology can be translated into stronger bounds on the parameter $b$.

 The model used for the accelerated bunch is a point particle of total charge $q=\,N \,e$ and total mass $m=\, N\, m_e$, where $N$ is the number of electrons accelerated with the effective acceleration, $m_e$ is the mass of a single electron and $e$ is its charge. Despite this is a very simplified model, it is still useful to extract qualitative phenomenological consequences of maximal acceleration in laser-plasma acceleration. Using this model, we will show that in the case of higher order jet electrodynamics, the existence of a maximal proper acceleration induces a damping in the {\it effective acceleration} reachable in laser-plasma acceleration which depends upon the {\it size} of the bunch being accelerated. For Born-Indeld theory, the maximal proper acceleration of the theory implies  an universal theoretical upper bound for the effective acceleration in laser-plasma acceleration systems.
\\
{\bf Notation.} In this paper $M$ is the $4$-dimensional spacetime manifold.
$TM$ is the tangent bundle of $M$, $J^k_0(M)$ is the $k$-order jet bundle and $\eta$ is a Lorenztian metric defined on $M$, although we will often assume that $\eta$ is the Minkowski metric with signature $(-1,1,1,1)$, since we will consider local properties of the spacetime. Local coordinates on $M$ are denoted by $(x^0,x^1,x^2,x^3)$ or simply by $x^{\mu}$ or similar notation. We will find useful  to identify the notation for points of $M$ and coordinates, since we are working on open domains of $M$ homeomorphic to $\mathbb{R}^4$. Natural coordinates on $TM$ are denoted by $(x^\mu,\dot{x}^\mu)$ with $\mu$ ranging from $0$ to $3$. In this work we have used natural units where the speed of light in vacuum is $c=1$. For a charged particle, $m$ will be the mass and $q$ the charge; $m_e$ is the mass of a single electron and $e$ its charge.

\section{Theories of electrodynamics with a maximal acceleration}
\subsection{Higher order jet electrodynamics}
Let us introduce first the theory of higher order jet electrodynamics discussed in \cite{Ricardo 2017}. The theory is based on the idea that, in order to embrace situations where the reaction of charged particles due to the interaction with classical fields must be taken into account, the electromagnetic field is described by a {\it generalized electromagnetic field} depending upon the differential higher order second jet associated to the charged test particle, as a section of a jet bundle. This is an example of {\it generalized field}, where the physical classical field depends on the way it is probed by test particles \cite{Ricardo2012}. Thus let us first assume that the generalized electromagnetic field has the form \cite{Ricardo 2017}
\begin{equation}
\bar{F}(\,^kx)=\bar{F}(x,\dot{x},\ddot{x},\dddot{x},..., x^{(k)})=\big(F_{\mu\nu}(x)+
\Upsilon_{\mu \nu}(x,\dot{x},\ddot{x},\dddot{x},..., x^{(k)})\big)d_4 x^{\mu}\wedge d_4 x^{\nu},
\label{electromagneticfield}
\end{equation}
 where $x:I\to M$ is the world line of the test particle and the dot-notation indicates derivatives respect to the proper time parameter of the theory. $d_4$ is the exterior differential operator on generalized forms, which is defined in very similar way as the standard exterior differential operator. Indeed, there is a well defined Cartan calculus for generalized fields, developed in close analogy with the classical Cartan calculus. The mathematical details for these constructions  can be found in \cite{Ricardo2012}. Also, albeit $\bar{F}$ is formulated in the general case of $k$-jet field, we will only need for our applications the case $k=2$.

In the theory of higher order jet electrodynamics, the spacetime structure that determines the proper time is a metric of maximal acceleration \cite{Ricardo2015}. This is a type of metric structure that depends up to the second derivatives of the world line coordinates (the second jet of the world line $x:I\to M$). In local coordinates, the metric of maximal acceleration is the bilinear form
\begin{align}
g:=\Big(1+ \frac{  \eta(D_{x'}x'(\tau),
D_{x'}x'(\tau))}{{a}^2_{max}\,\eta(x',x')}\Big)\,\eta.
\label{maximalaccelerationmetric0}
\end{align}
Here the $'$-notation denotes derivatives respect to the limit Lorentzian metric $\eta$ and $D$ is the associated covariant derivative. Proper acceleration $D_{\dot{x}}\dot{x}$ respect to the metric $g$ is bounded by $a_{max}$.

Under these assumptions and after some further constraints related with the conservation of energy by assuming compatibility with the covariant Larmor's radiation formula \cite{Jackson}, one obtains a model of the point charged particle described by a second order differential equation. The derivation of this equation can be found in \cite{Ricardo 2017}. In local coordinates, it is given by the relation
\begin{align}
m\,\ddot{x}^{\mu} =\,q\,F^{\mu}\,_{\nu}\,\dot{x}^{\nu}-\,\frac{2}{3}\,{q^2}\,g_{\rho\sigma}\,
\ddot{x}^{\rho}\ddot{x}^{\sigma}\,\dot{x}^{\mu},\quad F^{\mu}\,_{\nu}=\,g^{\mu\rho}F_{\rho\nu}.
\label{equationofmotion}
\end{align}
This is our proposed equation of motion for a classical point charged particle. Formally, it is similar to the Lorentz-Dirac equation \cite{Dirac}, but the Scott's term is absented.  Let us remark briefly that the underlying spacetime metric of the theory is the metric of maximal acceleration and that for such a metric the kinematical constraints for the speed and higher order derivatives are different than for a Lorentzian metric. In particular, $\dot{x}$ and $\ddot{x}$ are not necessarily orthogonal respect to the metric of maximal acceleration $g$ \cite{Ricardo 2017}.

Remarkably, the value of the maximal proper acceleration \eqref{valueofthemaximalacceleration} depends on the characteristics of charged particle, in particular, on the charge and mass of the particle. It is this property that we will show has potential phenomenological consequences for laser-plasma acceleration.

 In {\it equation} \eqref{equationofmotion}, if we contract the left hand side with the left hand side and the right hand side with the right hand side using the Minkowski metric $g$, then the following expression is obtained in a Fermi coordinate system for $\eta$,
 \begin{align*}
 m^2\,a^2\,= \,F^2_L -\,\left(\frac{2}{3}\,q^2\right)^2\,(a^2)^2\,\dot{x}^\mu\dot{x}_\mu.
 \end{align*}
 where
 \begin{align*}
 F^2_L :=\, q\, F_{\mu\nu}\, \dot{x}^\nu\,\eta^{\mu\rho}\,q\,F_{\rho\sigma}\dot{x}^\sigma\,\geq 0
 \end{align*}
  and where
  \begin{align*}
 a^2:=\,g(\ddot{x},\ddot{x}).
 \end{align*}
If $\eta(x',x')=-1$, then the relation $(2.2)$ is  Since
  \begin{align*}
  g=\,\left(1-\frac{\eta(x'',x'')}{a^2_{max}}\right)\,\eta.
  \end{align*}
As a consequence of that,
 \begin{align*}
 \eta(\ddot{x},\ddot{x})\geq \,g(\ddot{x},\ddot{x}).
 \end{align*}

From the equation of motion \eqref{equationofmotion}, one obtains the relation
\begin{align*}
m^2\,a^2 \,= &\,F^2_L \,+\left(\frac{2}{3}\,q^2\right)^2\,(a^2)^2\geq \left(\frac{2}{3}\,q^2\right)^2\,(a^2)^2.
\end{align*}
Since $F^2_L$ is positive, it follows the bound for the maximal acceleration
\begin{align}
g(\ddot{x},\ddot{x})\leq \left(\frac{3}{2}\,\frac{m}{q^2}\right)^2 =\,a^2_{\mathrm{max}}.
\label{valueofthemaximalacceleration}
\end{align}

The consequences from maximal acceleration that we will discuss later are tied to very large accelerations. Therefore, we need to understand the domain of validity of the equation \eqref{equationofmotion}. Although initially the theory was developed in the perturbative regime where $a^2<<a^2_{\mathrm{max}}$, it is shown in the {\it Appendix} that  if the following three conditions hold good, then the model is valid for accelerations close to the maximal acceleration:
\begin{enumerate}
\item The domain of validity of the metric structure \eqref{maximalaccelerationmetric0} is maximal, that is, valid in the open domain of $J^2_0(M)$ defined by the conditions $a^2<\,a^2_{\mathrm{max}}$ and $\vec{v}\cdot \vec{v}\leq 1$ (causal test particles).

\item The generalized covariant Larmor's law as presented in \cite{Ricardo 2017} holds good, but for all the physical situations of interest, it can be replaced by the standard covariant Larmor's power law.

\item A formal generalization of the renormalization of mass parameter is valid, such that non-perturbative terms originated for large accelerations contribute to the dress of the bare mass parameter to the physical mass parameter.

\end{enumerate}
As we will show in the {\it Appendix}, under these assumptions, equation \eqref{equationofmotion} can be applied in the domain when $a^2$ is close to $a^2_{\mathrm{max}}$.

Let us make some remarks on the above assumptions. We start with the second one. Since the equations of motion of charged equations can be obtained by energy-momentum balance arguments, it implies that in essence the model proposed here will violate the energy-momentum balance, since not all the power of the generalized electromagnetic field is taken into account when approximating the generalized Larmor's law by the standard covariant Larmor's law (see the final argument at the {\it Appendix}). However, let also us note that the use of a renormalization of mass procedure as indicated in point three is an sign that our account of the problem of the motion of a point charged particle can only have a phenomenological status. Then the apparent un-balance of energy-momentum is a consequence of the approximation to the electromagnetic field in the generalized Poynting vector leading to the generalized Larmor's power law. Therefore, in situations when the strength of the radiation reaction field is small compared with the external field we can postulate that the standard covariant Larmor's law is a good approximation, despite that the accelerations of the individual charge can be close to $a_\mathrm{max}$. The un-balance of energy-momentum is only an artifact of our considerations and it has a small impact on the dynamics.

 This generalization of the domain of application of equation \eqref{maximalaccelerationmetric0} is consistent with the domain of dependence of the metric $g$ in the jet bundle $J^2_0(M)$. Thus the above argument fix the value of $a_{\mathrm{max}}$
in the whole theoretical domain of $g$.

\subsection{Born-Infeld kinematics} The second example of classical theory of electrodynamics with a maximal acceleration that we will consider is the {\it kinematical theory},  motivated by Born-Infeld theory \cite{BornInfeld} and formulated in terms of a pseudo-complex extension of Minkowski geometry  \cite{Schuller}. Such extensions determine a metric, which is isometric to a Sasaki type metric defined on $TM$ \cite{Schuller}. We do not need to discuss the details of Schuller's theory here except that it implies the existence of a maximal acceleration whose theoretical value is given by the expression
 \begin{align}
\tilde{a}_{\mathrm{max}}=\,\frac{q}{m}\, b^{-1},
\label{Born Infeld maximal acceleration}
\end{align}
where $b$ is the coupling of the Born-Infeld theory, a constant independent of the particle \cite{BornInfeld} and where it has been assumed that the Lorentz force equation is still valid in Born-Infeld theory. We observe that this value of the maximal acceleration that a charged particle can reach has a different dependence on the mass $m$ and the charge $q$ than the expression \eqref{valueofthemaximalacceleration}, implying different consequences in laser-plasma acceleration, as it will be shown later. In particular, the maximal electric field in Born-Infeld theory is given in natural units by the expression
\begin{align}
\widetilde{E}_{\mathrm{max}}=\,b^{-1}.
\label{maximal electric field 2}
\end{align}

\section{Phenomenological consequences of maximal acceleration in laser-plasma acceleration}

The phenomena that we will consider in the context of two different classical electrodynamical theories with a maximal acceleration concerns  laser-plasma acceleration systems. For a general review of the fundamental theory of laser-plasma acceleration, see for instance \cite{Esarey et al.}. In such dynamical systems, sustained high accelerations  of order $10^{21}\,m/s^2$ on bunches of electrons are achieved (see for instance \cite{Litos}). Such accelerations are close to the order of the maximal effective acceleration reachable for some specific bunches of charged particles in the framework of higher order electrodynamics, as we will show below using a toy model for the bunch of particles.

We assume that the bunch of accelerated particles evolves as a sole particle of mass $m=\,N\, m_e$ and charge $q=\,N\,e$, where $N$ is the number of electrons in the bunch being effectively accelerated. This is a crude approximation, but it will help us to show the qualitative differences between different theories of classical electrodynamics with a maximal acceleration in the context of laser-plasma acceleration phenomenology. Note that the description of the bunch as a single point charged particle is consistent with the fact that the bunch of $N$ electrons evolves in a consistent way during the acceleration. Indeed, such a consistent time evolution is an ideal goal in particle acceleration physics where, for instance, the notion of \emph{reference trajectory} in beam dynamics is fundamental in the theory and in the modelling of the particle accelerators beams \cite{Wiedemann}. The model for the bunch used shares the same general idea of ideal beam.

 Note that although the dynamical effects due to the radiation produced by betatron motion of the electrons composing the bunch are relevant in laser-plasma acceleration and could induce very large transversal accelerations on each individual particle \cite{di Piazza et al.}, they must necessarily averaged out in the effective dynamics of the accelerated bunch, which is the acceleration which we are considering applies to the bunch of particles. That this must be the case follows from the difference in the scales of the associated accelerations. The radiation due to betatron motion of the electrons composing the bunch implies scales of accelerations that are associated with X-ray and  $\gamma$-ray radiation spectrum \cite{di Piazza et al.}, scales that can be even shorter than the average spacing between electrons. Such accelerations range in the scales $c^2/\lambda_r \, \sim \,10^{28}-10^{29}\, m/s^2$, where $\lambda_r$ is the wavelength of the radiation due to betatron motion. However, the maximal effective acceleration, the acceleration that can be identified with the acceleration of such bunch as a whole,  were of order $10^{21}\, m/s^2$. This means that effects due to betatron motion are effectively neutralized in the range of scales considered.

 When adopting the point charged model discussed above, thermal effects are disregarded. Although such an approximation implies a systematic error too, thermal effects cannot hide the qualitative differences implied by different theories with maximal acceleration. This is because thermal effects are of order $10$ to $40 \%$ in some relevant observables \cite{Esarey et al.}, but this systematic error cannot erase the effects of the scaling of effective maximal acceleration with the {\it size} $N$ of the bunch that we will discuss below.

\subsection{Consequences of higher order jet electrodynamics in laser-plasma acceleration} If we adopt the above approximate description of the bunch of particles, in higher order jet electrodynamics, the maximal acceleration achievable for the collective bunch of trapped electrons is given by the expression
\begin{align}
a_{\mathrm{max}}(N):=\,\frac{3}{2}\,\frac{m_e}{e^2}\,\frac{1}{N}.
\label{maximal acceleration on a bunch 1}
\end{align}
The expression $\frac{3}{2}\,\frac{m_e}{e^2}$ corresponds to the maximal acceleration of a single electron,
\begin{align*}
a_{\mathrm{max}}(1)=\,\frac{3}{2}\,\frac{m_e}{e^2}\,\sim\, 10^{32} m/s^2.
\end{align*}
Now we see that the {\it size} $N$ of the accelerated particles bunch dumps the maximal effective acceleration achievable by electrodynamical means by a factor $1/N$. For current wakefield plasma accelerator facilities one can reach accelerated bunches  with  $N\sim 10^{8}$  of the order $a\sim 10^{21} \,m/s^2$ \cite{Litos}, a figure close to the current bounds for the maximal acceleration for a bunch of that {\it size}, which is in this case of order $a_{\mathrm{max}}\,\sim 10^{24}\, m/s^2$. Although for these accelerations $a$ are  still three orders of magnitude below $a_{\mathrm{max}}$, we could expect to reach the regime $a\sim a_{\mathrm{max}}$ by increasing number of particles $N$ in the bunch.

\subsection{Constrains in Born-Infeld kinematics from laser-plasma acceleration} In this case, the maximal value of the acceleration is given by the expression \eqref{Born Infeld maximal acceleration}, which has a trivial dependence with $N$.

On the other hand, laser-plasma accelerator systems currently can  provide a lower bound for the {\it Born-Infeld parameter} $b$, that comes from the acceleration gradients of order $10^{22}\, m/s^2$. From the expression \eqref{Born Infeld maximal acceleration},
\begin{align}
b^{-1}\geq \frac{m_e}{e}\,10^{22}\, m/s^2 \geq 10^{11}\, N/C,
\label{bound on b}
\end{align}
which is similar to the bound imposed by Thomas's precession \cite{Schuller 2}. However, advances in the intensity and the acceleration that one could reach in laser-plasma acceleration lead us to think that the bound \eqref{bound on b} can be strengthened by increasing the effective acceleration of the bunches.

\section{Discussion}

We have discussed how laser-plasma acceleration phenomenology can be used to probe classical theories with a maximal acceleration in electrodynamics. However, there are several problematic assumptions in our argument that we need to discuss.

The first one concerns the model of a bunch a sole particle. There are several issues on this practice that we would like to expose here.
The model for the bunch of accelerated point charged particles that we have used is a simplified model where the bunch is considered as a sole particle of mass $m= \,N\,m_e$ and charge $q=\,N\,q_e$. Although this model is useful to extract qualitative consequences for each theory, the rawness of the model makes necessary a critical justification of it.  It is in this context that the nature of the dynamical systems under investigations, namely, accelerated bunch of by a wakefield laser produced in a plasma, enters in favour of the model, because it is expected in the beam quality parameters, including a reduction of current energy spread and increase of the intensity by increasing the number of particles per bunch. This implies that the model that we use for the bunch as a sole particle will be more adequate for future laser-plasma acceleration beams. In this context, the effective acceleration of any of the point particles composing the bunch is indeed the acceleration of the bunch. Therefore, when we apply our model of point particle to the bunch, it will provide the world line of an hypothetical particle representing the bunch as a whole and if the quality of the beam is very good, this is a good description for the forecast of when and where to find the charge in spacetime, despite that in reality the bunch is formed by individual charged particles, with individual world lines suffering of tremendous transverse acceleration due to betatron motion \cite{Corde et al.}.

The above justification is emphasized by the fact that the same model for the bunch as a point charged particle implies different qualitative results for each of the theories analyzed. In particular, the dependence on the number of electrons $N$ in the bunch is totally different for the standard classical electrodynamics (where there is no limitation on the maximal acceleration field), for Born-Infeld theory (where the maximal acceleration only depends on the ratio $q/m$) and for higher order jet electrodynamics (where the maximal acceleration depends on $q^2/m$).

There is a similarity between our model for the accelerated bunch of charged particles and the concept of {\it coherent acceleration} \cite{Veksler}. However, the physical conditions where each of these concepts apply are rather different. In the case of laser-plasma acceleration, the wavelength of the accelerated bunch is of order of the plasma wavelength, while for coherent acceleration as proposed in \cite{Veksler} the scale of the system being accelerated must be much shorter than the plasma wavelength.

An important limitation of the model of the bunch as a point charged particle is its intrinsic incapability to describe dynamical effect due to radiation by betatron motion. We saw that such effects are relevant for the effective acceleration of a bunch when  acceleration is in the range $10^{28}-10^{29}\, m/s^2$. When the effective acceleration of the bunch reaches experimentally such scales, our model cannot be a good description.

  There are some further issues on the general applicability of classical electrodynamic models to the systems under consideration. Current investigations indicate that quantum electrodynamical effects are of fundamental relevance in the dynamics of laser-plasma acceleration \cite{di Piazza et al.}. Quantum electrodynamics involves a {\it maximal acceleration} associated with the critical field in classical electrodynamics \cite{Schwinger1951, di Piazza et al.}. Such acceleration is associated to dynamical quantum systems whose asymptotic states contain only one charged particle. For a system composed by an electron, the critical acceleration is of order $a_{cr}\sim 10^{30}\, m/s^2$. However, in our model the bunch of accelerated particles are classical dynamical systems, with a well assigned world line and with $(q,m)$ parameters  $N$ times larger than $(e,m_e)$ for a single electron and where the effective acceleration (not the individual transversal accelerations of each electron) is much smaller than the critical acceleration. This implies for higher order jet electrodynamics an effective theoretical maximal proper acceleration much smaller than the classical acceleration $a_{cr}$ associated to the critical field $F_{cr}$.

 Let us also remark that recent experimental evidence for radiation reaction quantum effects on the individual electrons composing the bunch has been reported in \cite{Cole, Poder}. There is no incompatibility between those findings and our ideas, since the analysis done by Cole et al. and by Poder et al. was not directly applied to the dynamical system of laser-plasma acceleration, but to the scattering of highly accelerated electrons by a wakefield laser-plasma acceleration with another secondary scattering laser. Moreover, the analysis were limited to specific electrodynamical models, namely, Lorentz force, Landau-Lifshitz equation, semi-classical models based upon Landau-Lifshitz and Quantum Electrodynamics. It could be very interesting to extend the considerations discussed in \cite{Cole} and also in \cite{Poder} to other classical and semi-classical models of charged particle, apart from the Landau-Lifshitz equation and Lorentz equation already discussed.

Finally, let us remark that the expression of the value for maximal acceleration coincidences with the value proposed by Caldirola's theory \cite{Caldirola} in the context of a very different theory of the electron: a finite difference equation where time is quantized in terms of a basic unit called the {\it chronon} and the maximal speed is the speed of light in vacuum $c$ \cite{Caldirola1956}. Under such hypotheses, the maximal acceleration in Caldirola's theory of the classical electron is also given by the expression
\begin{align*}
a_{\mathrm{max}}=\,\frac{3}{2}\,\frac{q^2}{m}.
\end{align*}
Therefore, our proposal to measure maximal acceleration in laser-plasma acceleration applies also to test Caldirola's theory, under the assumption that the nature of the chronon is not universal, but depends on the characteristics of the dynamical system. In addition, the fact that the same maximal acceleration is obtained in Caldirola's theory supports our argument that indeed \eqref{valueofthemaximalacceleration} is a non-perturbative result. On the other hand it also serve us to evidence that Caldirola's theory can be tested by the methodology discussed in this work.

 \appendix

\section{Derivation of the equation of motion for a point charged particle in higher order jet electrodynamics}
We provide in this {\it Appendix} a self-contained derivation of the equation of motion \eqref{equationofmotion}. The reader can follow several of the calculational details omitted here in \cite{Ricardo2015,Ricardo 2017}. The main difference with the derivation presented here and the derivation presented in \cite{Ricardo 2017} is that we keep all the higher order terms in $\epsilon =\,\frac{\eta(x'',x'')}{a^2_{\mathrm{max}}}$ in this new analysis, albeit at an abstract level. From one side, such higher order terms contribute to the generalized Larmor's law, but higher order derivative terms only contribute to the renormalization of the bare mass.
\subsection{Generalized higher order jet electrodynamics}
The generalized electromagnetic field is of the form $\bar{F}=\,F+\Upsilon$ such that
\begin{align}
\Upsilon\approx \,\epsilon^\alpha \,f(x)+\,\textrm{higher order terms in $\epsilon$},
\label{approximationupsilon}
\end{align}
where $\alpha$ is a positive constant. They are solutions of a formal generalization of Maxwell equations to higher order jet fields \cite{Ricardo 2017}.
Let us assume that the physical trajectory of a point charged particle is a smooth curve of class $\mathcal{C}^k$
such that $g(\dot{x},\dot{x})=-1$, with $\dot{x}^0<0$ and such that the square of the proper acceleration
$a^2$ is bounded from above. We have the following general expression for the generalization electromagnetic field \eqref{electromagneticfield},
\begin{align*}
\Upsilon_{\mu\nu}(x,\,\dot{x},\,\ddot{x},\,\dddot{x},...)=
\,B_{\mu}\dot{x}_{\nu}\,-B_{\nu}\dot{x}_{\mu}\,+
C_{\mu}\ddot{x}_{\nu}\,-C_{\nu}\ddot{x}_{\mu}\,
+D_{\mu}\dddot{x}_{\nu}\,-D_{\nu}\dddot{x}_{\mu}+...,
\end{align*}
with $\dot{x}_{\mu}=\,g_{\mu\nu}\,\dot{x}^{\nu}$, holds good.

The general form of the $k$-jet fields  $B(\tau),C(\tau),D(\tau)$  along the smooth curve $x:\mathbb{R}\to M$ are
\begin{align*}
& B^{\mu}(\tau)=\,\beta_1\,\dot{x}^{\mu}(\tau)\,+\,\beta_2\,\ddot{x}^{\mu}(\tau)\,+
\,\beta_3\,\dddot{x}^{\mu}(\tau)\,+\,\beta_4\,\ddddot{x}^{\mu}(\tau)+\cdot\cdot\cdot,\\
& C^{\mu}(\tau)=\,\gamma_1\,\dot{x}^{\mu}(\tau)\,+\,\gamma_2\,\ddot{x}^{\mu}(\tau)\,+
\,\gamma_3\,\dddot{x}^{\mu}(\tau)\,+\,\gamma_4\,\ddddot{x}^{\mu}(\tau)+\cdot\cdot\cdot,\\
& D^{\mu}(\tau)=\,\delta_1\,\dot{x}^{\mu}(\tau)\,+\,\delta_2\,\ddot{x}^{\mu}(\tau)\,+
\,\delta_3\,\dddot{x}^{\mu}(\tau)\,+\,\delta_4\,\ddddot{x}^{\mu}(\tau)+\cdot\cdot\cdot.
\end{align*}
If one expects that a general relativistic view is possible, then field $\bar{F}$ and the metric of maximal acceleration should have the same dependence, that is up to second order derivatives. Therefore, all higher derivatives terms must cancel in the above expressions. In particular, the following conditions
\begin{align}
\gamma_k\,=\delta_k\,=0,\, k\geq 0,\quad \beta_k=0, k\geq 3
\label{choice of the coefficients}
\end{align}
are shown to be sufficient for our purposes to have a second order differential equation for a point charged particle compatible with the generalized Larmor's law.

\subsection{Kinematical relations}
The kinematical relations for the metric $g$ are the following \cite{Ricardo 2017}. First, we recall the relation
 \begin{align*}
 g(\dot{x},\dot{x})=-1\quad \Leftrightarrow\quad \eta(x',x')=-1.
 \end{align*}
This is a particular case of the following relation
 \begin{align*}
 \eta(x',x')& =\,(1-\epsilon)^{-1}\,g(x',x')=\,(1-\epsilon)^{-1}\,g((1-\epsilon)^{1/2} \dot{x},(1-\epsilon )^{1/2}\dot{x})=\,g(\dot{x},\dot{x}).
 \end{align*}
Then the following kinematical constrains  \cite{Ricardo 2017} hold good,
\begin{align}
& g(\dot{x},\dot{x})=-1,\label{covariantkineticconstrain1}\\
& g(\dot{x},\ddot{x})=\,-\frac{\dot{\epsilon}}{2}+\,\mathcal{O}(\epsilon^2_0) \label{covariantkineticconstrain2}.
\end{align}
Here the error term $\mathcal{O}(\epsilon^2_0)$ represents higher derivatives and contributions in terms of power of $\epsilon$ and its time derivatives. $\epsilon_0$ was a measure of $\epsilon$ along compact interval of times, but the measure must be extended to derivatives of $\epsilon$ as well.

\subsection{Larmor's relativistic radiation law in  higher order jet electrodynamics}
Since the theory of higher order electrodynamics is formally analogous to Maxwell's theory,  it is natural to assume that by a similar procedure than in the standard theory, the relation
\begin{align}
\frac{d{p}^{\mu}_{rad}}{d\tau}=-\,\left( 1+c_1\,\epsilon^\alpha+\cdot\cdot\cdot\right) \frac{2}{3}\,q^2 (\ddot{x}^{\rho}\,\ddot{x}^{\sigma}g_{\rho\sigma})(\tau)\,\dot{x}^{\mu}(\tau)
\label{generalizedLarmor}
\end{align}
holds good.
$c_1$ is a constant, which is necessarily not large, since standard relativistic Larmor's law is experimentally very well established when $a<<\,\frac{3}{2}\,\frac{q^2}{m}$.

\subsection{Derivation of the new equation of motion of a point charged particle}

We assume a form of generalized Lorentz equation, where instead of a Maxwell field we have a generalized Maxwell field as a source of a generalized Lorentz force. Using the previous formulae for the generalized Faraday $2$-form $\bar{F}$, the {\it bare Lorentz force equation} is of the form
\begin{align*}
m_b(\tau)\,\ddot{x}^{\mu} & =\,q\,\bar{F}^{\mu}_\nu\,\dot{x}^\nu=\,q\,F^{\mu}\,_{\nu}\,\dot{x}^\nu+ \big(B^{\mu}\dot{x}_{\nu}-\,\dot{x}^{\mu}B_{\nu}\big)\dot{x}^{\nu}\\
& +\big(C^{\mu}\ddot{x}_{\nu}-\,\ddot{x}^{\mu}C_{\nu}\big)\dot{x}^{\nu} +\big(D^{\mu}\dddot{x}_{\nu}-\,\dddot{x}^{\mu}D_{\nu}\big)\dot{x}^{\nu}\,+...,
\end{align*}
 with $F^{\mu}\,_{\nu}=\,g^{\mu\rho}F_{\rho\sigma}.$ $m_b(\tau)$ is the bare mass parameter. In principle, $m_b(\tau)$ depends on $\tau$-time parameter.
 On the right hand side of the above expression all the contractions that appear in expressions as
 $\big(B^{\mu}\dot{x}_{\nu}-\,\dot{x}^{\mu}B_{\nu}\big)\dot{x}^{\nu}$, etc,  are performed with the metric $g$.

With the choice \eqref{choice of the coefficients}, one gets the relation
\begin{align}
m_b(\tau)\,\ddot{x}^\mu =\, q\,F^\mu\,_\nu\,\dot{x}^\nu\,+\beta_2\,\ddot{x}^\mu\,\dot{x}^\nu\,\dot{x}_\nu-\,\dot{x}^\mu \,
\beta_2\,\ddot{x}_\nu\,\dot{x}^\nu.
\end{align}
Using the kinetic relations \eqref{covariantkineticconstrain1} and \eqref{covariantkineticconstrain2} for $g$, one obtains the expression
\begin{align}
m_b(\tau)\,\ddot{x}^{\mu} =\,q\,F^{\mu}\,_{\nu}\dot{x}^\nu-\beta_2 \ddot{x}^{\mu}-\beta_2\,\left(
-\frac{1}{2}\dot{\epsilon}+\,\textrm{higher order terms}\right)\dot{x}^{\mu},
\label{equationbeforerenormalization}
\end{align}
where here the higher order terms stand for the higher order in $\epsilon$ and its derivatives that appear in the contraction $\ddot{x}^\nu\dot{x}_\nu$ using the metric $g$.
This relation must be consistent with the generalized covariant Larmor's power radiation formula \eqref{generalizedLarmor}. Therefore, we require
\begin{align}
&\beta_2=\, \left( 1+c_1\,\epsilon^\alpha+\cdot\cdot\cdot\right)\frac{4}{3}q^2\,\ddot{x}^\nu\,\ddot{x}_\nu\,\left(
-\frac{1}{2}\dot{\epsilon}+\,\textrm{higher order terms}\right)^{-1}.
\end{align}
In the remaining correction terms in equation \eqref{equationbeforerenormalization}, only powers on the acceleration $\ddot{x}$ or velocity vector $\dot{x}$ appear.

The next step is to consider the re-normalization of the bare mass in the following way,
\begin{align}
 m_b(\tau)+\,\frac{2}{3}\,q^2\,\ddot{x}^\nu\,\ddot{x}_\nu\,\left(
-\frac{1}{2}\dot{\epsilon}+\,\textrm{higher order terms}\right)^{-1}=\,m,
\label{renormalizationofmass}
\end{align}
a well defined expression, except in a set of measure zero. The mass parameter $m$ does not depend on the time parameter.
Then we obtain the following differential constraint for a point charged particle,
\begin{align}
m\,\ddot{x}^{\mu} =\,q\,F^\mu\,_\nu\dot{x}^\nu\,-\frac{2}{3}\,q^2\,\left( 1+c_1\,\epsilon^\alpha+\cdot\cdot\cdot\right)\,\ddot{x}^\nu\,\ddot{x}_\nu\,\dot{x}^{\mu}.
\label{prerenormalization}
\end{align}
Under the assumption that the standard covariant Larmor's law is valid, one finds that the equation
\begin{align*}
m\,\ddot{x}^{\mu} =\,q\,F^{\mu}\,_{\nu}\,\dot{x}^{\nu}-
\,\frac{2}{3}\,{q^2}\,\ddot{x}^{\nu}\ddot{x}_{\nu}\,\dot{x}^{\mu},\quad F^{\mu}\,_{\nu}=\,g^{\mu\rho}F_{\rho\nu}
\label{equationofmotion}
\end{align*}
holds good almost everywhere in the domain of definition of the theory. We do not consider here the exceptions or singularities of the equation, that have been introduce through the different normalizations and renormalizations procedures. The case of the singularities requires individual treatment, in analogy to the case $\dot{\epsilon}=0$ study in \cite{Ricardo 2017}. We think that similar methods can be follow in this new treatment, but we post-pone the treatment for future investigations.

 We have seen that, as long us the standard covariant Larmor's law is valid and the other assumptions of the theory hold almost everywhere, at least at a phenomenological level, the equation of motion of a point charged particle in higher order jet electrodynamics is the equation \eqref{equationofmotion} almost everywhere in the domain of definition of the theory in $J^2_0(M)$.

We remark here that the approximation from the generalized covariant Larmor's law to the standard covariant Larmor's law is justified as long as the magnitude of the radiation field emitted by the bunch is small compared with the wakefield. Furthermore, if the approximation  on the power radiation law is not assumed, then the corresponding equation of motion will remain a second order ODE, since in the extension $\Upsilon$ of $F$, only second order jets are involved.

 \subsection*{Acknowledgements} I would like to acknowledge to the Frankfurt Institute for Advanced Studies for support during part of  the development of this paper.
\small{
}

\end{document}